\documentclass[12pt]{article}
\usepackage{amsbsy,amsmath}
\usepackage{amsfonts}
\usepackage{amssymb}
\usepackage{amscd}
\usepackage{bbm}
\usepackage{fancybox}
\usepackage{cite}
\usepackage{amsmath,amsfonts,amsbsy}
\usepackage{pstricks,pst-node}
\usepackage[small,bf,hang]{caption2}
\usepackage[linktocpage]{hyperref}
\usepackage{graphicx}
\usepackage{epsfig}
\usepackage{psfrag}
\usepackage{comment}
\usepackage{xcolor}
\usepackage[normalem]{ulem}

\psset{unit=1.3cm,linewidth=.5pt,radius=.2}  

\usepackage{multirow}                     
\usepackage{float}                          
\usepackage{lscape}                         
\usepackage{bm}

\newcommand{\beq}{\begin{equation}}
\newcommand{\eeq}{\end{equation}}
\newcommand{\bi}{\begin{itemize}}
\newcommand{\ei}{\end{itemize}}
\newcommand{\bt}{\begin{tabular}}
\newcommand{\et}{\end{tabular}}
\newcommand{\bc}{\begin{center}}
\newcommand{\ec}{\end{center}}

\newcommand{\be}{\begin{equation}}
\newcommand{\ee}{\end{equation}}
\newcommand{\bea}{\begin{eqnarray}}
\newcommand{\eea}{\end{eqnarray}}
\newcommand{\ba}{\begin{array}}
\newcommand{\ea}{\end{array}}

\def\bbox{{\,\lower0.9pt\vbox{\hrule \hbox{\vrule height 0.2 cm
\hskip 0.2 cm \vrule height 0.2 cm}\hrule}\,}}
\newcommand{\dsl}{\pa \kern-0.5em /}

\newcommand{\LL }{\mathcal{L}}
\newcommand{\TT }{\mathcal{T}}

\makeatletter \@addtoreset{equation}{section} \makeatother

\def\slashchar#1{\setbox0=\hbox{$#1$}           
   \dimen0=\wd0                                 
   \setbox1=\hbox{/} \dimen1=\wd1               
   \ifdim\dimen0>\dimen1                        
      \rlap{\hbox to \dimen0{\hfil/\hfil}}      
      #1                                        
   \else                                        
      \rlap{\hbox to \dimen1{\hfil$#1$\hfil}}   
      /                                         
   \fi}

\begin{document}

\begin{titlepage}
\begin{center}

\vskip 1.5cm

{\Large  Causality and Energy Conditions in Nonlinear Electrodynamics}

\vskip 1cm

{\bf Jorge G. Russo${}^{1,2}$ and Paul K.~Townsend${}^3$} \\

\vskip 25pt

{\em $^1$  \hskip -.1truecm
\em Instituci\'o Catalana de Recerca i Estudis Avan\c{c}ats (ICREA),\\
Pg. Lluis Companys, 23, 08010 Barcelona, Spain, \vskip 5pt}

{email: {\tt jorge.russo@icrea.cat}} 

\vskip .4truecm

{\em $^2$  \hskip -.1truecm
\em Departament de F\' \i sica Cu\' antica i Astrof\'\i sica and Institut de Ci\`encies del Cosmos,\\ 
Universitat de Barcelona, Mart\'i Franqu\`es, 1, 08028
Barcelona, Spain. \vskip 5pt}

\vskip .4truecm

{\em $^3$ \hskip -.1truecm
\em  Department of Applied Mathematics and Theoretical Physics,\\ Centre for Mathematical Sciences, University of Cambridge,\\
Wilberforce Road, Cambridge, CB3 0WA, U.K. \vskip 5pt }

{email: {\tt P.K.Townsend@damtp.cam.ac.uk}} \\

\vskip .4truecm

\end{center}

\vskip 0.5cm
\begin{center} {\bf ABSTRACT}\\[3ex]
\end{center}

For the general theory of nonlinear electrodynamics (NLED), we prove that causality implies both the Dominant Energy Condition (DEC) and, surprisingly, the Strong Energy Condition (SEC). This has implications for gravitational applications, such as regular black holes supported by NLED matter. For self-dual NLED theories, weak-field causality alone 
implies both the DEC and SEC, as we illustrate with Born-Infeld and ModMax electrodynamics.

\bigskip

\vfill

\end{titlepage}
\tableofcontents

\section{Introduction}

A standard framework for nonlinear electrodynamics (NLED) is a Lagrangian density function $\LL(S,P)$, where $(S,P)$ are the 
two Lorentz (pseudo)scalars quadratic in the Faraday field-strength tensor \cite{Born:1933qff,Born:1934gh,Boillat:1970gw,Plebanski:1970zz,Bialynicki-Birula:1984daz}. 
In the context of a 4D spacetime 
with metric ${\rm g}$ (of ``mostly plus'' signature), 
\begin{equation}
S = - \frac14 {\rm g}^{\mu\rho}{\rm g}^{\nu\sigma} F_{\mu\nu} F_{\rho\sigma}\, , \qquad 
P = -\frac{1}{8 \sqrt{|{\rm g}|}}\varepsilon^{\mu\nu\rho\sigma} F_{\mu\nu}F_{\rho\sigma}\, , 
\end{equation}
where $|{\rm g}| = -\det{\rm g}$. Following \cite{Schellstede:2016zue} we shall say that the NLED theories of this framework 
constitute the  ``Plebanski class''.  Limits of Plebanski-class NLED theories may be outside this class; several physical examples are known \cite{Bialynicki-Birula:1984daz,Russo:2022qvz,Mezincescu:2023zny} but they have no weak-field limit and will not be considered here. 

The stress-energy tensor for the generic NLED in the Plebanski class can be found from the Hilbert formula
\begin{equation}
\TT_{\mu\nu} = - \frac{2}{\sqrt{|{\rm g}|}} \frac{\partial (\sqrt{|{\rm g}|}\LL)}{\partial {\rm g}^{\mu\nu} } \, . 
\end{equation} 
For example, for $\LL=S$ (Maxwell) the stress-energy tensor is
\begin{equation}
\TT^{\rm Max}_{\mu\nu} ={\rm g}^{\rho\sigma} F_{\mu\rho}F_{\nu\sigma } + S{\rm g}_{\mu\nu}\, , 
\end{equation}
which is traceless and satisfies\footnote{This follows from the identity $X^4 =2SX^2 + P^2 \mathbb{I}$ for $4\times 4$ matrix $X$ with entries $X_\mu{}^\nu = F_{\mu\lambda} {\rm g}^{\lambda\nu}$.}
\begin{equation}\label{T2Max} 
 {\rm g}^{\rho\sigma} \TT^{\rm Max}_{\mu\rho}\TT^{\rm Max}_{\nu\sigma} 
= (S^2+P^2){\rm g}_{\mu\nu}\, . 
\end{equation} 
More generally, using the relations
\begin{equation}
\frac{\partial S}{\partial {\rm g}^{\mu\nu}} = \frac12\left(S  {\rm g}_{\mu\nu} - \TT_{\mu\nu}^{\rm Max}\right)\, , \qquad 
\frac{\partial P}{\partial {\rm g}^{\mu\nu}}  = \frac12 P {\rm g}_{\mu\nu}\, , 
\end{equation}
we find that 
\begin{equation}\label{STform}
\TT_{\mu\nu} = \LL_S \TT_{\mu\nu}^{\rm Max} + (\LL - S\LL_S -P\LL_P) {\rm g}_{\mu\nu} \, ,
\end{equation}
where $(\LL_S,\LL_P)$ are the partial derivatives of $\LL(S,P)$. All scalar functions of the stress-energy tensor can be written as functions of the following two independent
Lorentz scalars:
\bea
\label{cinco}
&& \Theta := {\rm g}^{\mu\nu} \TT_{\mu\nu} = 4 (\LL - S\LL_S -P\LL_P)\, \ ,
\\
&&\TT^2:={\rm g}^{\mu\nu} {\rm g}^{\rho\sigma}\TT_{\mu\rho}\TT_{\nu\sigma} =4(\LL-S\LL_S-P\LL_P)^2+4(S^2+P^2)\LL_S^2\ .
\label{seis}
\eea

One aim of this paper is  to determine how conditions on the stress-energy tensor, such as the Dominant Energy Condition (DEC) and the Strong Energy Condition (SEC),  depend on the choice of  function $\LL(S,P)$. We assume that  
\be\label{vacE} 
\LL(0,0)= 0\, ,  
\ee
in which case the DEC and SEC are satisfied in the NLED vacuum. We then use our results for the DEC and SEC constraints 
on $\LL(S,P)$ to investigate the relationship between energy conditions and causality conditions. Our principal result, applicable to all NLED theories in the Plebanski class, is  that causality implies both the DEC and the SEC. As the DEC implies 
the Weak Energy Condition (WEC) and both the WEC and the SEC imply the Null Energy Condition (NEC), this means that 
{\sl any NLED theory (in the Plebanski class) that violates any of these four energy conditions is acausal}. 

The fact that any causal NLED theory must satisfy the DEC is easily proved from convexity conditions, which are 
among the necessary and sufficient conditions for causality found in \cite{Schellstede:2016zue}; we call them the ``weak-field'' causality conditions because they can be violated, generically, for weak fields\footnote{Weak in relation to the scale 
introduced by interactions in generic NLED theories.}. 
However, it should be appreciated that the DEC is not sufficient, generically, even for weak-field causality. 

The proof that the SEC is also required by causality is more involved because it is a consequence (generically) of one further ``strong-field'' causality condition 
\cite{Schellstede:2016zue}. An alternative derivation of this causality condition
was given in \cite{Russo:2024kto} along with a precise criterion for the weak-field/strong-field distinction. More recently, it was noticed in \cite{Russo:2024llm} that the strong-field causality condition becomes much simpler when $\LL$ is viewed as a function of the 
new variables
\be\label{UV}
V= \frac12\left(\sqrt{S^2+P^2} + S\right) \, , \qquad 
U= \frac12\left( \sqrt{S^2+P^2} - S\right)\, . 
\ee
Equivalently,
\be
S= V-U\, , \qquad P^2 = 4UV\, . 
\ee
Notice that both $U$ and $V$ are non-negative parity-even scalars. If we restrict to NLED theories that preserve parity
then $\LL(S,P)$ can depend on $P$ only through $P^2$, and we can then trade $(S,P^2)$ for $(U,V)$; i.e. we may replace $\LL(S,P)$ by $\LL(U,V)$. With this assumption, we prove that
strong-field causality implies the SEC. 

For the special class of self-dual NLED theories (those with a Hamiltonian invariant under a $U(1)$ electromagnetic-duality group) strong-field causality is implied by weak-field causality \cite{Russo:2024llm} from which it follows that weak-field causality alone implies both 
the DEC and the SEC. We illustrate this result with the examples of Born-Infeld \cite{Born:1934gh} and its ModMax-type generalization \cite{Bandos:2020jsw,Bandos:2020hgy} that we call ``ModMaxBorn''. 

We conclude with a discussion of some implications for applications of NLED theories to black-hole physics, and we make contact with recent work on ``stress-energy flows'' in the NLED context.

\section{Energy Conditions}\label{sec:EC}

The task of determining the conditions on $\LL(S,P)$ required for the stress-energy tensor to satisfy the DEC, SEC
or other energy conditions is greatly simplified by the formula \eqref{STform} because $\mathcal{L}_S>0$ (for 
reasons detailed in \cite{Schellstede:2016zue}) and the Maxwell stress-energy tensor satisfies both the DEC and the 
SEC (and hence the WEC and NEC). This means that the first term in \eqref{STform} satisfies all of the 
above-mentioned energy conditions. We only need to investigate the effect of the second term in \eqref{STform} on 
the DEC and SEC. We now turn to a study of these two energy conditions. 

\subsection{Dominant Energy Condition}

The DEC requires
\begin{equation}
\xi^\mu\zeta^\nu \TT_{\mu\nu} \ge0   
\end{equation}
for any pair of timelike vector fields $(\xi,\zeta)$ with the same time orientation.
From \eqref{STform}, we see that this imposes the condition
\begin{equation}
\LL_S \left(\xi^\mu\zeta^\nu \TT_{\mu\nu}^{\rm Max}\right) + 
(\LL - S\LL_S -P\LL_P) (\xi\cdot \zeta) \ge 0\, . 
\end{equation}
The first term is non-negative. Since $\xi\cdot \zeta <0$, the second term is also non-negative provided that
\begin{equation}\label{convex1}
S\LL_S + P\LL_P \ge \LL   \, . 
\end{equation}
This equation is satisfied whenever $\LL$ is a convex function of $(S,P)$ satisfying \eqref{vacE}.
This follows from the fact that a convex function lies on or above all its tangents planes, which is
equivalent, for all $(S,P)$ and $(S_0,P_0)$, to
\begin{equation}
(S-S_0)\LL_S + (P-P_0)\LL_P \ge \LL(S,P) - \LL(S_0,P_0)\, . 
\end{equation}
Choosing $S_0=P_0=0$, and assuming that $\LL(0,0)=0$ we arrive 
at \eqref{convex1}. 

For a twice-differentiable function, an equivalent definition of a convex function is one with a positive Hessian 
matrix (no negative eigenvalues). Thus, if $\LL(S,P)$ is convex we have 
\begin{equation}\label{convexity}
\LL_{SS} + \LL_{PP} \ge0 \, , \qquad \LL_{SS}\LL_{PP} - \LL_{SP}^2 \ge0\, .  
\end{equation}
These inequalities imply that $\LL_{SS}$ and $\LL_{PP}$ are {\sl separately} non-negative. As shown in 
\cite{Bandos:2021rqy} they are equivalent to the conditions for convexity of $\LL$ {\sl viewed as a function of the 
electric field}, which guarantees an equivalent Hamiltonian formulation. They are also the conditions required for 
(weak-field) causality \cite{Schellstede:2016zue}.  Thus, the stress-energy tensor of any causal NLED satisfies the DEC. 

\subsection{Strong Energy Condition}

The SEC is the statement that 
\begin{equation}\label{SEC}
\xi^\mu\xi^\nu \left( \TT_{\mu\nu} - \frac12 \Theta {\rm g}_{\mu\nu}\right) \ge0 \, ,  
 \end{equation}
for any timelike vector field $\xi$, which we may normalize such that $\xi^2=-1$.  For the stress-energy tensor of \eqref{STform},  this equality becomes
\begin{equation}\label{SECineq}
\LL_S \left(\xi^\mu\xi^\nu \TT^{\rm Max}_{\mu\nu}\right) +  (\LL - S\LL_S -P\LL_P)\ge  0\, . 
\end{equation}
As the first term is non-negative but the second term is non-positive for any causal theory (we used this to prove the DEC)
a detailed examination of both terms is needed to determine whether the SEC is satisfied or violated. 

At this point we shall assume that parity is preserved since this allows us to view $\LL$ as a function of the variables $(U,V)$ 
defined in  \eqref{UV}. We then have 
\be\label{LS}
\LL_S  = \frac{V\LL_V -U\LL_U}{V+U}\, ,  \qquad S\LL_S + P\LL_P = V\LL_V + U\LL_U\, .
\ee
In addition, we may choose  local coordinates $\{t,{\bf x}\}$ in a neighbourhood of any point in the spacetime, and a Lorentz frame at that point for which $\xi= \partial_t$. The SEC inequality of \eqref{SECineq} then reduces to 
\be 
K\left(V\LL_V -U\LL_U\right) + \left(\LL -U\LL_U- V\LL_V\right)  \ge0 \, , \qquad K := \frac{\TT^{\rm Max}_{tt}}{V+U} \, .
\ee
We may rewrite this as the inequality 
\be\label{SEC-K}
\LL - 2U\LL_U \ge  \left(1-K\right) \left(V\LL_V - U\LL_U\right)\, . 
\ee
Using the expressions 
\be\label{express}
\TT^{\rm Max}_{tt} = \frac12\left(|{\bf E}|^2 + |{\bf B}|^2\right)\, , \qquad 
V+U = \frac12 \sqrt{\left(|{\bf E}|^2 + |{\bf B}|^2\right)^2- 4|{\bf E}\times{\bf B}|^2}\, , 
\ee
where $({\bf E},{\bf B})$ are the electric and magnetic field components of the field-strength tensor $F$, we
see that $K\ge1$, with equality iff $|{\bf E}\times{\bf B}|=0$. The right-hand side of \eqref{SEC-K} is 
therefore non-positive, and zero for non-vacuum field configurations only when $K=1$. 
As $(U,V)= (|{\bf B}|^2,|{\bf E}|^2)$ for $K=1$, which does not restrict the range of values of $(U,V)$,
the SEC inequality can be satisfied for all $({\bf E},{\bf B})$ iff
\be\label{SECfinal}
C(U,V) := \LL(U,V) -2U\LL_U \ge0 
\ee
for all $(U,V)$  

It is not difficult to show that \eqref{SECfinal} is true for $U=0$; i.e. on the positive $V$-axis
that is part of the boundary of the positive quadrant in the $(U,V)$-plane. It follows immediately from 
\be
C(0,0) = \LL(0,0) =0\, , \qquad C_V(0,V) = \LL_V(0,V) = \LL_S|_{U=0}>0\, , 
\ee
where we have used the expression for $\LL_S$ in \eqref{LS}. This shows that $C(0,V)$ is a 
positive function of $V$ for all positive $V$. If we can show that $C(U,V)$ is an increasing 
function of $U$ at fixed non-negative $V$ (i.e, $C_U(U,V)>0$ for $V>0$) then $C(U,V)$ will be positive for all points in the positive $(U,V)$ quadrant (in the domain of $\LL$), and the SEC will be satisfied. 

From the definition of $C(U,V)$ in \eqref{SECfinal} we have
\be
C_U= -\left(\LL_U + 2U\LL_{UU}\right)\, ,   
\ee
but the strong-field causality condition derived in \cite{Schellstede:2016zue,Russo:2024kto} was shown in \cite{Russo:2024llm} to be equivalent to 
\begin{equation}\label{dos}
\LL_U+2U \LL_{UU} <0\, ,   
\end{equation}
which we now see to be equivalent to the inequality $C_U>0$. 

To summarise, the strong-field causality condition \eqref{dos} is equivalent (on the assumption that  $\LL$ is zero in the vacuum) to the inequality 
\be
\LL(U,V) -2U \mathcal{L}_U \ge0\, , 
\ee
with equality only in the vacuum ($U=V=0$), and this implies the SEC.

Notice, however, that $C_U>0$ is not a necessary condition for the SEC, since $C(U,V)$ could still be positive in the $(U,V)$ quadrant if $C(U,V)$ is a sufficiently slowly {\sl decreasing} function of $U$ for positive $V$. As we shall see in section \ref{sec:ex} the original Born theory is an example of an NLED theory for which the SEC is satisfied despite the fact that the strong-field causality condition is violated. Nevertheless, physical NLED theories must be causal, and hence they must also satisfy both the DEC and the SEC.

 \section{Self-Dual NLED}\label{sec:sd}
 
 The condition for  $\LL(U,V)$ to define a self-dual NLED  is the simple partial differential equation $\LL_U\LL_V=-1$ \cite{Gibbons:1995cv}; its solution for initial conditions $\LL(0,V) =\ell(V)$ can be found in  \cite{C&H}: 
\begin{equation}\label{C&H}
\LL = \ell(\tau) -\frac{2U}{\dot\ell(\tau)} \, , \qquad \tau= V + \frac{U}{\dot\ell^2(\tau)}\, ,  \end{equation}
where $\dot\ell(\tau)= d\ell(\tau)/d\tau$. It follows from \eqref{C&H} that 
\be\label{LULV}
\LL_V = \dot \ell\, , \qquad \LL_U= -1/\dot\ell\, , 
\ee
and hence that $\LL_U\LL_V=-1$. 

The equation for $\tau$ implies that $\tau\ge0$ with 
equality only in the vacuum, and the assumption that $\LL(0,0)=0$ becomes 
\be 
\ell(0)=0\, . 
\ee 
The equation for $\LL_S$ in \eqref{LS} now becomes
\be
\LL_S = \frac{\dot\ell^2 V+U}{\dot\ell(V+U)}\, , 
\ee
and hence $\LL_S>0$ becomes 
\be\label{simpineq}
\dot\ell >0\, . 
\ee
The stress-energy tensor of \eqref{STform} may now be written as 
\begin{equation}\label{sd-ST}
\TT_{\mu\nu} = \left[\frac{\tau\dot\ell}{U+V}\right] \TT^{\rm Max}_{\mu\nu} +
(\ell - \tau \dot\ell) {\rm g}_{\mu\nu} \, .  
\end{equation}
For self-dual theories, the two independent Lorentz scalars \eqref{cinco} and \eqref{seis} become
\begin{equation}   
\Theta = 4(\ell - \tau\dot\ell)\, ,   
\qquad
 \TT^2  =  4\left[(\tau\dot\ell)^2 + (\ell -\tau\dot\ell)^2\right] \, .  
 \label{T2}
\end{equation}
Notice that these expressions for $\Theta$ and $\TT^2$ are both functions of $\tau$ only, either explicitly or implicitly via the 
function $\ell(\tau)$.  As all Lorentz invariants constructed from the stress tensor can be expressed as functions of $\Theta$ and $\TT^2$, they are also all expressible as functions of $\tau$.  This feature has implications that we discuss briefly 
in the conclusions.

We move now to implications for energy conditions

\subsection{SEC and DEC}

We have seen that the SEC inequality is equivalent to \eqref{SECfinal}. Using \eqref{C&H} and \eqref{LULV} we may rewrite this expression as the remarkably simple inequality
\be
\ell(\tau) \ge0 \, , 
\ee
but this is an immediate consequence of \eqref{simpineq}, given $\ell(0)=0$.

For the stress-energy tensor of \eqref{sd-ST} the DEC is the inequality  
 \be
 \left[\frac{\tau\dot\ell}{U+V}\right] \left(\xi^\mu\zeta^\nu  \TT^{\rm Max}_{\mu\nu}\right) + (\ell - \tau\dot\ell) (\xi\cdot\zeta) \ge0\, , 
 \ee
 for any two timelike vector fields with the same time orientation. As the first term is non-negative and $\xi\cdot\zeta <0$, the DEC
 will be satisfied if 
 \be\label{sd-DEC}
 \ell(\tau) - \tau\dot\ell(\tau)  \le0 
 \ee
 for all non-negative $\tau$ in the domain of the function $\ell$.
 
 The convexity/causality conditions 
 of \eqref{convexity} reduce for all self-dual NLED (with a weak-field limit) to the very simple inequalities \cite{Russo:2024llm}
\begin{equation}\label{ell-ineqs}
\dot\ell \ge 1\, , \qquad \ddot\ell\ge 0\, .
\end{equation}
Only the second of these inequalities is relevant to the DEC; it implies that $\ell(\tau)$ is a convex function, and hence (for $\tau_0<\tau$) that
\begin{equation}
\int_{\tau_0}^\tau \left[\dot\ell(\tau)-\dot\ell(\tau- \zeta)\right] d\zeta \ge0\, , 
\end{equation}
which is equivalent to 
\begin{equation}\label{first-order}
(\tau-\tau_0) \dot\ell(\tau) - [\ell(\tau) - \ell(\tau_0)] \ge 0 \, . 
\end{equation}
This is the statement that the graph of the function $\ell$ lies on or above 
all its tangent lines, which is another definition of a convex function. Setting 
$\tau_0=0$ we get precisely the inequality of \eqref{sd-DEC}. 

As the inequality $\dot\ell\ge1$ was not needed, the DEC is, even for self-dual NLED, insufficient for even weak-field causality. Nevertheless, we confirm that all causal self-dual 
NLED theories necessarily satisfy both the DEC and the SEC.

\section{Examples}\label{sec:ex}

We shall illustrate our results with a few examples. We first consider the Born-Infeld theory, defined by  \cite{Born:1934gh}, 
\begin{equation}\label{LLBI}
\LL_{\rm BI} = T- \sqrt{T^2-2TS -P^2}\, ,  
\end{equation}
and the original Born theory, with \cite{Born:1933qff}
\be\label{OBorn}
\LL_{\rm Born} = T- \sqrt{T^2-2TS}\, .  
\end{equation}
Whereas Born-Infeld is a self-dual and causal theory, the original Born theory is neither self-dual nor causal; it fails to satisfy the strong-field causality condition \cite{Schellstede:2016zue}. Indeed, recalling that $S=V-U$, we have
$$
(\LL_U+2U\LL_{UU})_{\rm Born}=-\frac{T^{\frac12}(T-2V)}{(T+2U-2V)^{\frac32}}\ ,
$$
which becomes positive in the interval $T<2V<T+2U$, violating \eqref{dos}.  

As we shall see, both the Born and Born-Infeld theories satisfy both the DEC and the SEC. For Born-Infeld this is a consequence of causality. In contrast, the Born theory satisfies the 
SEC despite the fact that it violates the strong-field causality condition (which is, of course, an essential condition for a physical theory). This further illustrates the point made in some detail in \cite{Russo:2024kto} that it is not easy to find causal NLED theories that are not self-dual; even the DEC and SEC are insufficient to guarantee causality. 

We then consider ModMax, which is a one-parameter family of conformal self-dual NLED theories 
parametrized by a coupling constant $\gamma$ \cite{Bandos:2020jsw}; for $\gamma\ge0$ the theory is causal, with Maxwell being the $\gamma=0$ case. A related example is the ModMax-type generalization of the Born-Infeld theory proposed in \cite{Bandos:2020jsw,Bandos:2020hgy} that we refer to as ``ModMaxBorn''.  

\subsection{Born and Born-Infeld}

The Born-Infeld theory is an example of a self-dual theory, found from the choice \cite{Russo:2024llm}
\be\label{ell-BI}
\ell(\tau) = T- \sqrt{T(T-2\tau)} \, ,  
\ee
where $T$ is the positive Born parameter with dimensions of energy density and the constant term is included to ensure that 
$\ell(0)=0$. Since 
\be
\dot\ell(\tau) = \frac{T^\frac12}{ (T- 2\tau)^\frac12}  \ge1\, , \qquad \ddot\ell(\tau) = \frac{T^\frac12}{(T- 2\tau)^\frac32} >0\, , 
\ee
we confirm that Born-Infeld is causal. The DEC is an immediate consequence of the fact that $\ell(\tau)$ is convex, and the SEC
is an immediate consequence of the fact that $\ell(\tau)$ is non-negative. 

For the purposes of pedagogy, we shall now verify that the BI stress-energy tensor satisfies both the DEC and the SEC by returning to the DEC and SEC inequalities used in section \ref{sec:EC}. A calculation yields 
\begin{equation}\label{BIcase}
\left[S\LL_S + P\LL_P  -\LL\right]_{\rm BI} = \LL_S(\LL_{\rm BI}-S)\, .
\end{equation}
Using this in \eqref{STform}, we find that
\begin{equation}
\TT^{\rm BI}_{\mu\nu} = \LL_S \left[\TT_{\mu\nu}^{\rm Max } + (S- \LL_{\rm BI}) {\rm g}_{\mu\nu}\right]\, . 
\end{equation}
Since $\LL_S>0$, the DEC is equivalent to 
\be\label{DECequiv}
\left(\xi^\mu\zeta^\nu \TT_{\mu\nu}^{\rm Max} \right) + (S- \LL_{\rm BI})(\xi\cdot\zeta)\ge0\, , 
\ee
for any pair of timelike vector fields with the same time orientation. The first term is positive, as is the second term if
$\LL_{\rm BI}>S$, which is true, so the DEC is satisfied. For Born's original theory, the
inequality \eqref{DECequiv} still applies, but with $\LL_{\rm BI}\to\LL_{\rm Born}$, and because
$\LL_{\rm Born} >S$ the DEC is again satisfied. 

Now we consider the SEC. The SEC inequality \eqref{SEC} reduces for BI to 
\begin{equation}\label{SEC-BI} 
(i_\xi F)^2 + \LL_{\rm BI}\,  \xi^2\ge 0\, , 
\end{equation} 
for all timelike $\xi$, which we may may normalise by requiring $\xi^2=-1$.  Choosing a Lorentz 
frame for which $\xi= \partial_t$, the inequality \eqref{SEC-BI} reduces to $|{\bf E}|^2 \ge \LL_{\rm BI}$, which is equivalent to 
\begin{equation}\label{SECBI2}
\sqrt{(T-|{\bf E}|^2)(T+|{\bf B}|^2) + |{\bf E}\times{\bf B}|^2} \ge T-|{\bf E}|^2\, . 
\end{equation}
Reality of the left hand side requires the ``Born bound''
 \begin{equation}
|{\bf E}|^2 \le T + \frac{|{\bf E}\times{\bf B}|^2}{T + |{\bf B}|^2}\, . 
 \end{equation}
Only those field configurations satisfying this bound are in the domain of $\LL_{\rm BI}$ . 
Notice that non-zero $|{\bf E}\times{\bf B}|$ allows $|{\bf E}|^2>T$, in which case the SEC inequality \eqref{SECBI2}
 is trivially satisfied. For $|{\bf E}|^2\le T$ both sides of \eqref{SECBI2} are non-negative,  so we can square them to arrive at the equivalent inequality 
\begin{equation}
|{\bf E}|^2 \le T + \frac{|{\bf E}\times{\bf B}|^2}{|{\bf E}|^2 + |{\bf B}|^2}\, ,  
 \end{equation}
 which is trivially satisfied for $|{\bf E}|^2\le T$. Thus, the BI stress-energy tensor satisfies the SEC in addition to the DEC. 

For the original Born Theory of \eqref{OBorn} the SEC inequality analogous to \eqref{SECBI2}
is
\be\label{BORNSEC}
\sqrt{T^2-T |{\bf E}|^2 + T|{\bf B}|^2} \ge T-|{\bf E}|^2\, . 
\end{equation}
The Born bound is now $|{\bf E}|^2 \le T+ |{\bf B}|^2$. This allows $|{\bf E}|^2\ge T$ but then 
\eqref{BORNSEC} is trivially satisfied. Therefore we need consider only $|{\bf E}|^2<T$, which allows us to take the square on both sides of \eqref{BORNSEC}; this yields
\be
|{\bf E}|^2 (T-|{\bf E}|^2) \ge -T |{\bf B}|^2\, , 
\ee
which is satisfied for $|{\bf E}|^2<T$. The SEC is therefore satisfied. 

\subsubsection{ModMax and ModMaxBorn}

The $\ell$ function for ModMax is \cite{Russo:2024llm}
\be\label{ell-MM}
\ell(\tau) = e^\gamma \tau \qquad (\gamma\ge0).
\ee
In this case 
\be
\dot \ell = e^\gamma \ge1 \, , \qquad \ddot\ell=0\, ,  
\ee
which confirms that ModMax is causal. The DEC is satisfied because $\ell$ is a convex function of $\tau$ (although not ``strictly convex'' because its Hessian is zero). The SEC is satisfied because $\ell(\tau)$ is positive for all positive $\tau$. Notice that these conclusions concerning the DEC and SEC remain true for $\gamma<0$, although the NLED theory is then acausal. This demonstrates that the combined DEC and SEC conditions do not imply causality, whereas both are implied by causality.  

The Born-Infeld extension of ModMax, that we call ModMaxBorn for sake of brevity, has the 
following $\ell$-function \cite{Russo:2024llm}. 
\be
\ell(\tau) = T- \sqrt{T(T-2e^\gamma \tau)}  \qquad (\gamma \ge0).
\ee
Notice that this reduces to \eqref{ell-BI} for $\gamma=0$, and to \eqref{ell-MM} in the (weak-field) $T\to\infty$ limit. 
We now have
\be
\dot\ell(\tau) = \frac{e^\gamma T^\frac12}{ (T- 2e^\gamma\tau)^\frac12}  \ge1\, , \qquad 
\ddot\ell(\tau) = \frac{e^{2\gamma}T^\frac12}{(T- 2e^\gamma\tau)^\frac32} >0\, , 
\ee
which tells us that ModMaxBorn is causal.   It also tells us (since $\ell$ is a convex function)  that the stress-energy tensor
satisfies the DEC. The SEC is also satisfied because $\ell(\tau) \ge0$ for non-negative $\tau$ in the domain of $\ell$ (which
imposes the upper bound $\tau\le \frac12 e^{-\gamma} T$).  Notice that the $\dot\ell\ge1$ condition is satisfied only for 
$\gamma\ge0$; the fact that it is violated for $\gamma<0$ tells us that causality requires $\gamma\ge0$ (in accord with what we already know from its limit to ModMax). For $\gamma<0$ we therefore have an acausal NLED theory for which the DEC and SEC are
satisfied. Thus,  causality is again a stronger condition (generically) than either the DEC or SEC, separately or combined.




\section{Discussion}

Over the past few decades, there have been many applications of nonlinear 
electrodynamics to gravitational physics, notably black holes (but also cosmology), in which many 
results rely on assumptions, explicit or implicit,  about energy conditions. The only fundamental
physical constraint on NLED theories in the Plebanski class (i.e. defined by a Lagrangian density 
that is a scalar function of the electric and magnetic fields alone, without derivatives) is causality, 
so the only fundamental constraints on the stress-energy tensor of a generic NLED theory are those
required by causality. Our main result is that causality implies that the NLED stress-energy tensor satisfies both the Dominant Energy 
Condition (DEC) and the Strong Energy Condition (SEC), but neither of these energy conditions, 
separately or combined, is sufficient to guarantee causality. 

More specifically, we have shown that weak-field causality implies the DEC whereas strong-field 
causality implies the SEC. The strong-field causality condition was
originally found (as recently as 2016) as a rather complicated inequality involving first and 
second derivatives of the Lagrangian density \cite{Schellstede:2016zue}. Its simplification to \eqref{dos} found in \cite{Russo:2024llm} is what made possible the conclusions of this paper concerning its relation to the SEC, which came as a surprise to us because we know of no previous connection of the SEC to causality. This result provides some welcome further intuition into the physical meaning of the strong-field causality condition.  

We should again emphasise that all these results are based on the premise that the NLED vacuum
has zero energy density, and hence that any violation of the DEC or SEC is possible only for non-vacuum field configurations. Adding a positive (negative) constant to the zero vacuum energy will not affect  the DEC (SEC) but will lead to a violation of the SEC (DEC) in the vacuum. As this added constant energy density has no effect on the NLED field equations, and is only relevant once gravity is included, it makes sense to consider it as a contribution to the cosmological constant.
 
 If the cosmological constant is zero then there is a Minkowski vacuum spacetime, and energy can be defined in General Relativity (GR) relative to the zero energy of an asymptotic Minkowski vacuum. In this context, 
 positive energy requires the matter stress-energy tensor to satisfy the Dominant Energy Condition (DEC). 
 We have shown here that the DEC is satisfied by any causal NLED theory, so the energy of any asymptotically 
 flat solution of  any NLED-Einstein field equations generalising the Maxwell-Einstein equations will be positive
 provided that the NLED theory is causal. 
 
 The SEC is generally considered as less fundamental than the DEC, but it is needed for the 
 theorem of Hawking and Penrose showing that  black holes formed in gravitational collapse must have a singularity
 in their interior\footnote{An earlier weaker theorem of Penrose assumes the NEC which is implied by the DEC; we refer the reader to \cite{Wald:1984rg} for details of these singularity theorems.}.  
 This means that the stress-energy tensor supporting any regular black hole 
 (see e.g. \cite{Borde:1996df,Hayward:2005gi,Lemos:2011dq})
must violate the SEC somewhere in the interior (see e.g. \cite{Zaslavskii:2010qz}
for a discussion of this point). Our results therefore imply that the Einstein field equations sourced by the stress-energy tensor of any causal NLED theory (with zero vacuum energy) cannot have a regular black hole solution. This has implications for the many claims in the literature (e.g. \cite{Ayon-Beato:1998hmi,Bronnikov:2000vy,Dymnikova:2004zc,Balart:2014cga,dePaula:2023ozi}) that certain  NLED-Einstein theories have regular black hole solutions\footnote{Typically, the regular ``core'' replacing a Reissner-Nordstrom-type singularity supports a de Sitter metric, and hence violates the SEC.}.

 Finally, we return to the observation in section \eqref{sec:sd} that all Lorentz scalar functions of the stress-energy tensor are functions of the independent variable $\tau$ of the $\ell$-function associated to a self-dual NLED. This allows us to make contact with the currently active topic of deformations of a given self-dual NLED theory by some Lorentz scalar function of its stress-energy tensor, which is necessarily a function of the trace $\Theta$
 of the $\TT$-tensor and the quadratic scalar $\TT^2$ constructed from it. 
 The idea is that a one parameter deformation leads to a `flow' from the initial theory to some other `target' self-dual NLED theory \cite{Ferko:2023wyi}.  As the space of self-dual theories is parameterised by one-variable functions 
$\ell(\tau)$, any such `flow' must correspond to a one-parameter family of $\ell$-functions. 
 As mentioned in section \ref{sec:sd}, both $\Theta$ and $\TT^2$ are functions of $\tau$ 
 only, and this implies a simple relation between the deformation of
 $\ell(\tau)$ and the deformation of the corresponding Lagrangian density.
 
For simplicity, let us consider a flow, to first order in a parameter $\epsilon$, from the free-field Maxwell theory; this corresponds to an $\ell$-function of the form 
\be
\ell_\epsilon(\tau) = \tau + \epsilon f(\tau) + \mathcal{O}(\epsilon^2)\, . 
\ee
Using \eqref{C&H}, we find that the corresponding Lagrangian density function is
\be
\LL_\epsilon(U,V) = V-U + \epsilon f(V+U) + \mathcal{O}(\epsilon^2).
\ee
Equivalently, since $V-U= \LL$ and $V+U=\tau$, for the Maxwell case, 
\be
\LL_\epsilon = \LL_{\rm Max} + \epsilon f(\tau_{\rm Max})\, . 
\ee
In other words, to first order in $\epsilon$, the deformation functions for  $\ell_{\rm Max}$ and $\LL_{\rm Max}$ are identical; the only difference is that $\tau$ is an independent variable in one case and a dependent variable in the other case. Moreover, from \eqref{T2} we have\footnote{The ``Max'' suffix here refers to ``Maxwell'' and should not be confused with the maximum value of $\tau$ in particular NLED theories, such as Born-Infeld.} 
\be
\tau^2_{\rm Max} = \frac14 \TT^2_{\rm Max} \, . 
\ee
We thus get a $\TT^2$-deformation of $\LL_{\rm Max}$ when $f(\tau)\sim \tau^2$. Consider $f= \frac12\tau^2$; since $\epsilon$ now has dimensions of inverse energy density we set $\epsilon=1/T$, where $T$ is a Born-type parameter, to get 
\be 
\ell(\tau) = \tau  + \frac{\tau^2}{2T}  + \mathcal{O}(1/T^2)\, , 
\label{exell}
\ee
which is the expansion of $\ell_{\rm BI}(\tau)$ to first order in $1/T$. 
This yields the expansion of $\LL_{\rm BI}$ to first order in $1/T$, which we can write as
\be
\LL = \LL_{\rm Max} + \frac{1}{8T} \TT^2_{\rm Max} + \mathcal{O}(1/T^2)\, . 
\label{grrs}
\ee
Of course, any other function $\ell(\tau)$ of the form \eqref{exell} will yield \eqref{grrs}; 
in order to get the BI theory unambiguously one needs the full expansion of 
$\ell_{\rm BI} (\tau)$. In general, the expansion of the function $\ell(\tau) $ to all orders in $\tau $ contains the data needed to specify any `flow' from Maxwell to any other self-dual NLED `target' theory.

\section*{Acknowledgements}
PKT has been partially supported by STFC consolidated grant ST/T000694/1. 
JGR acknowledges financial support from grants 2021-SGR-249 (Generalitat de Catalunya) and MINECO  PID2019-105614GB-C21.




\begin{thebibliography}{10}


\bibitem{Born:1933qff}
M.~Born,
``Modified field equations with a finite radius of the electron,''
Nature \textbf{132} (1933) no.3329, 282.1

\bibitem{Born:1934gh}
M.~Born and L.~Infeld,
``Foundations of the new field theory,''
Proc. Roy. Soc. Lond. A \textbf{144} (1934) no.852, 425-451

\bibitem{Boillat:1970gw}
G.~Boillat,
``Nonlinear electrodynamics - Lagrangians and equations of motion,''
J. Math. Phys. \textbf{11} (1970) no.3, 941-951

\bibitem{Plebanski:1970zz}
J.~Plebanski, ``Lectures on non-linear electrodynamics'', (The Niels Bohr Institute and NORDITA, Copenhagen, 1970).

\bibitem{Bialynicki-Birula:1984daz}
I.~Bialynicki-Birula,
``Nonlinear Electrodynamics: Variations on a theme by Born and Infeld'',
in {\sl Quantum Theory of Particles and Fields}, eds. B. Jancewicz and
J. Lukierski, (World Scientific, 1983) pp. 31-48.

\bibitem{Schellstede:2016zue}
G.~O.~Schellstede, V.~Perlick and C.~L\"ammerzahl,
``On causality in nonlinear vacuum electrodynamics of the Pleba\'nski class,''
Annalen Phys. \textbf{528}, no.9-10, 738-749 (2016)
[arXiv:1604.02545 [gr-qc]].



\bibitem{Russo:2022qvz}
J.~G.~Russo and P.~K.~Townsend,
``Nonlinear electrodynamics without birefringence,''
JHEP \textbf{01} (2023), 039
[arXiv:2211.10689 [hep-th]].

\bibitem{Mezincescu:2023zny}
L.~Mezincescu, J.~G.~Russo and P.~K.~Townsend,
``Hamiltonian birefringence and Born-Infeld limits,''
JHEP \textbf{02}, 186 (2024)
[arXiv:2311.04278 [hep-th]].



\bibitem{Russo:2024kto}
J.~G.~Russo and P.~K.~Townsend,
``Born Again,''
[arXiv:2401.04167 [hep-th]].

\bibitem{Russo:2024llm}
J.~G.~Russo and P.~K.~Townsend,
``On Causal Self-Dual Electrodynamics,''
[arXiv:2401.06707 [hep-th]].




\bibitem{Bandos:2020jsw}
I.~Bandos, K.~Lechner, D.~Sorokin and P.~K.~Townsend,
``A non-linear duality-invariant conformal extension of Maxwell's equations,''
Phys. Rev. D \textbf{102}, 121703 (2020)
[arXiv:2007.09092 [hep-th]].

\bibitem{Bandos:2020hgy}
I.~Bandos, K.~Lechner, D.~Sorokin and P.~K.~Townsend,
``On p-form gauge theories and their conformal limits,''
JHEP \textbf{03}, 022 (2021)
[arXiv:2012.09286 [hep-th]].

\bibitem{Bandos:2021rqy}
I.~Bandos, K.~Lechner, D.~Sorokin and P.~K.~Townsend,
``ModMax meets Susy,''
JHEP \textbf{10}, 031 (2021)
[arXiv:2106.07547 [hep-th]].


\bibitem{Gibbons:1995cv}
G.~W.~Gibbons and D.~A.~Rasheed,
``Electric - magnetic duality rotations in nonlinear electrodynamics,''
Nucl. Phys. B \textbf{454}, 185-206 (1995)
[arXiv:hep-th/9506035 [hep-th]].


\bibitem{C&H}
R. Courant and D. Hilbert, ``Methods of Mathematical Physics'', Vol.II  (Wiley Interscience, 1962) pp.91-94. 

\bibitem{Wald:1984rg}
R.~M.~Wald,
``General Relativity,''
Chicago Univ. Pr., 1984,

\bibitem{Borde:1996df}
A.~Borde,
``Regular black holes and topology change,''
Phys. Rev. D \textbf{55}, 7615-7617 (1997)
[arXiv:gr-qc/9612057 [gr-qc]].
  
\bibitem{Hayward:2005gi}
S.~A.~Hayward,
``Formation and evaporation of regular black holes,''
Phys. Rev. Lett. \textbf{96}, 031103 (2006)
[arXiv:gr-qc/0506126 [gr-qc]].

\bibitem{Lemos:2011dq}
J.~P.~S.~Lemos and V.~T.~Zanchin,
``Regular black holes: Electrically charged solutions, Reissner-Nordstr\"om outside a de Sitter core,''
Phys. Rev. D \textbf{83}, 124005 (2011)
doi:10.1103/PhysRevD.83.124005
[arXiv:1104.4790 [gr-qc]].


\bibitem{Zaslavskii:2010qz}
O.~B.~Zaslavskii,
``Regular black holes and energy conditions,''
Phys. Lett. B \textbf{688} (2010), 278-280
[arXiv:1004.2362 [gr-qc]].



\bibitem{Ayon-Beato:1998hmi}
E.~Ayon-Beato and A.~Garcia,
``Regular black hole in general relativity coupled to nonlinear electrodynamics,''
Phys. Rev. Lett. \textbf{80}, 5056-5059 (1998)
[arXiv:gr-qc/9911046 [gr-qc]].

\bibitem{Bronnikov:2000vy}
K.~A.~Bronnikov,
``Regular magnetic black holes and monopoles from nonlinear electrodynamics,''
Phys. Rev. D \textbf{63} (2001), 044005
[arXiv:gr-qc/0006014 [gr-qc]].

\bibitem{Dymnikova:2004zc}
I.~Dymnikova, 
``Regular electrically charged structures in nonlinear electrodynamics coupled to general relativity,''
Class. Quant. Grav. \textbf{21} (2004), 4417-4429
[arXiv:gr-qc/0407072 [gr-qc]].

\bibitem{Balart:2014cga}
L.~Balart and E.~C.~Vagenas,
``Regular black holes with a nonlinear electrodynamics source,''
Phys. Rev. D \textbf{90} (2014) no.12, 124045
[arXiv:1408.0306 [gr-qc]].

\bibitem{dePaula:2023ozi}
M.~A.~A.~de Paula, H.~C.~D.~Lima, Junior, P.~V.~P.~Cunha and L.~C.~B.~Crispino,
``Electrically charged regular black holes in nonlinear electrodynamics: Light rings, shadows, and gravitational lensing,''
Phys. Rev. D \textbf{108} (2023) no.8, 084029
[arXiv:2305.04776 [gr-qc]].




\bibitem{Ferko:2023wyi}
C.~Ferko, S.~M.~Kuzenko, L.~Smith and G.~Tartaglino-Mazzucchelli,
``Duality-invariant nonlinear electrodynamics and stress tensor flows,''
Phys. Rev. D \textbf{108}, no.10, 106021 (2023)
[arXiv:2309.04253 [hep-th]].



 
\end{thebibliography}
\end{document}